\documentclass[a4paper,fleqn]{cas-sc}
\pdfoutput=1
\usepackage[numbers]{natbib}
\usepackage{amsmath,amssymb,amsfonts}
\usepackage{algorithmic}
\usepackage{graphicx}
\usepackage{textcomp}
\usepackage{multirow}
\usepackage{caption}
\usepackage{subcaption}
\usepackage{setspace}

\def\tsc#1{\csdef{#1}{\textsc{\lowercase{#1}}\xspace}}
\tsc{WGM}
\tsc{QE}
\tsc{EP}
\tsc{PMS}
\tsc{BEC}
\tsc{DE}

\begin{document}

\let\WriteBookmarks\relax
\def\floatpagepagefraction{1}
\def\textpagefraction{.001}

\shorttitle{Quantifying the Impact of Data Characteristics on the Transferability of Sleep Stage Scoring Models}

\shortauthors{A. Supratak and P. Haddawy}

\title [mode = title]{Quantifying the Impact of Data Characteristics on the Transferability of Sleep Stage Scoring Models}




%
\author[1]{Akara Supratak}[type=editor,
                        auid=000,bioid=1,
                        orcid=0000-0002-6739-7642]

\ead{akara.sup@mahidol.edu}



\author[1,2]{Peter Haddawy}[type=editor,
                        auid=001,bioid=2,
                        orcid=0000-0003-2203-006X]

\cormark[1]

\ead{peter.had@mahidol.ac.th}




\address[1]{Faculty of ICT, Mahidol University, 999 Phuttamonthon 4 Road Nakhon Pathom, 73170, Thailand}
\address[2]{Bremen Spatial Cognition Center, University of Bremen, Bremen, Germany}

\cortext[cor1]{Corresponding author}



\begin{abstract}
Deep learning models for scoring sleep stages based on single-channel EEG have been proposed as a promising method for remote sleep monitoring. However, applying these models to new datasets, particularly from wearable devices, raises two questions.
First, when annotations on a target dataset are unavailable, which different data characteristics affect the sleep stage scoring performance the most and by how much?
Second, when annotations are available, which dataset should be used as the source of transfer learning to optimize performance?
In this paper, we propose a novel method for computationally quantifying the impact of different data characteristics on the transferability of deep learning models.
Quantification is accomplished by training and evaluating two models with significant architectural differences, TinySleepNet and U-Time, under various transfer configurations in which the source and target datasets have different recording channels, recording environments, and subject conditions.
For the first question, the environment had the highest impact on sleep stage scoring performance, with performance degrading by over 14\% when sleep annotations were unavailable.
For the second question, the most useful transfer sources for TinySleepNet and the U-Time models were MASS-SS1 and ISRUC-SG1, containing a high percentage of N1 (the rarest sleep stage) relative to the others.
The frontal and central EEGs were preferred for TinySleepNet.
The proposed approach enables full utilization of existing sleep datasets for training and planning model transfer to maximize the sleep stage scoring performance on a target problem when sleep annotations are limited or unavailable, supporting the realization of remote sleep monitoring.
\end{abstract}


\begin{highlights}
\item Fully-computational approach to quantify transferability in sleep scoring
\item Impact of characteristics quantified via relative performance differences
\item Transferability quantified via fine-tuning models under various settings
\item Different recording environments had the highest impact on model performance

\end{highlights}


\maketitle

\doublespacing

\section{Introduction}
\label{sec:introduction}

Sleep is vital in promoting mental and physical health~\cite{irwin2015}.
The ability to monitor how well we sleep can alert us when the quality of our sleep declines, indicating possible need for medical attention.
Such monitoring requires admitting subjects to sleep clinics to run a  polysomnography (PSG) sleep diagnosis, which involves collecting and analyzing multiple biosignals such as electroencephalogram (EEG), electrooculogram (EOG), electromyogram (EMG), and electrocardiogram (ECG).
However, this approach is expensive and time-consuming, and does not scale well to a larger population due to the limited number of sleep experts, who will be scoring sleep, and the limited number of beds in the clinics.
It would be preferable to monitor sleep quality remotely, which could provide an early warning to seek expert medical advice.

Recently, many studies have proposed remote single-channel sleep monitoring to address such problems EEG~\cite{supratak2017,supratak2020,mousavi2019,phan2019,back2019,sun2019,fiorillo2021}. Single-channel signals can be collected from wearable devices such as  ear-EEGs~\cite{tabar2020,mikkelsen2019} and a headbands~\cite{koushik2019}.
Deep learning has gained popularity for scoring sleep stages based on single-channel EEG due to its ability to extract useful features from raw signals, thereby eliminating the need for hand-crafted features.
Typically, convolutional neural networks (CNNs) have been used in the top layers to transform raw PSG epochs into useful features and recurrent neural networks (RNNs) have been used in the bottom layers to learn temporal information such as transition rules~\cite{back2019,mousavi2019,supratak2017,supratak2020,kuo2021,nie2021recsleppnet}.
Several studies have proposed transforming raw PSG epochs into spectral-based images, typically small ones, before training deep learning models~\cite{phan2019,elmoaqet2022}.
 Recently, a fusion model that combines DeepSleepNet~\cite{supratak2017} and SeqSleepNet~\cite{phan2019} has been introduced to leverage features from both raw single-channel EEG and its spectral-based representation for sleep stage classification~\cite{wang2022}.
In addition, some researchers have suggested reducing the number of parameters in deep learning models by replacing the stack of convolutional layers with time-frequency domain feature extraction algorithms, while still using a bi-directional LSTM to learn sleep stage transition rules~\cite{you2022}.
The majority of this work, however, involves training deep learning models from scratch, which requires large amounts of labeled sleep data, and such data can be very expensive to collect.

When a labeled sleep dataset for the task at hand is not available, a possible solution is to directly apply a model that has been pre-trained on a publicly available sleep dataset to the target problem.
The challenge here is that when applying a pre-trained model to a new set of subjects, there may be differences in environment, hardware configurations and/or subject conditions (e.g., healthy or with sleep disorders).
This raises the research question: \textit{(RQ1) Which different data characteristics between the public dataset and the target problem affect the model performance the most and by how much?}
An answer to this can help estimate how well a pre-trained model would perform in a practical application.
Alternatively, we may have some data for the target problem, but not sufficient data to train a model from scratch. 
In this case, a possible solution is to utilize a transfer learning technique to fine-tune a pre-trained model on the target dataset~\cite{phan2019transfer,radha2021}. 
Even though it has been shown that such transfer learning can help improve the performance on the target dataset, there is still the research question: \textit{(RQ2) Which dataset should be used as the source of transfer to maximize the performance on the target dataset?}
One could try transferring models pre-trained on all existing datasets to a target dataset, but this is computationally expensive due to the high cost of training. Therefore, a method that can predict transferability at a much lower computational cost is needed.

A recent study proposed a computational approach to establish the relationship among computer vision tasks~\cite{zamir2018}.
The authors demonstrated that such a relationship can be used to suggest a better source for transfer learning (i.e., a better starting point) compared to a random choice when training a model for a specific task.
Such estimation of the source’s applicability to a particular task or dataset is commonly referred to as transferability.
Inspired by the limitations of this work that require model fine-tuning on the target dataset, several studies have proposed statistical and information-theoretic methods to predict the suitability of a transfer source without any fine-tuning ~\cite{tran2019,bao2019,nguyen2020}.

If we could quantify the similarity among sleep datasets with different characteristics as \cite{zamir2018}, we could determine which sleep dataset is a better transfer source than the others and plan an effective model transfer.
This would be the first step in sleep stage scoring that could lead to future improvements in predicting transferability without the need for model fine-tuning.
To the best of our knowledge, this is the first to study to quantify the similarity among sleep datasets with different data characteristics and to evaluate the relation to the transferability of deep learning models trained on them.

In this study, we aim to answer the two research questions RQ1 and RQ2 to determine in advance which different data characteristics should be avoided and which source dataset is most useful (i.e., has a higher performance gain) for a target problem.
The main contributions of this study are as follows:
\begin{itemize}
    \item For RQ1, we propose a new means of quantifying the impact that various differences in data characteristics between public datasets and target datasets have on the sleep stage scoring performance of deep learning models.
    A number of publicly available sleep datasets are utilized to simulate different characteristics such as recording channels, recording environments, and subject conditions.
    \item For RQ2, we investigate the characteristics of source datasets that lead to good transferrability performance.  We do this by a new means of quantifying the transferability of deep learning models based on a fully-computational approach that leverages the performance differences obtained from various transfer learning configurations.
    Such quantification generates a collection of transferability scores of the deep learning models pre-trained on different sleep datasets.
    This collection can be utilized to determine which source datasets would be useful for a given target dataset (e.g., wearable devices) and to plan model transfer.
    To the best of our knowledge, such relations in sleep stage scoring have not been investigated to date.
\end{itemize}

This study takes a different approach to sleep stage scoring from other studies that have developed transfer learning techniques~\cite{phan2019toward}.
Other transfer learning studies aim to utilize a large sleep dataset to maximize the model performance on several target datasets.
In contrast, we aim to understand the impact of various characteristics of sleep datasets on the direct applicability of pre-trained models, as well as their transferability to new datasets. 
This enables us to fully utilize the existing sleep datasets for training and planning model transfer to maximize the sleep stage scoring performance on a target problem when sleep annotations are limited or unavailable.
The findings of this study contribute to the realization of remote sleep monitoring from home environments.

\section{Materials and Methods}

\subsection{Problem Setup}
\label{sec:method_prob_setup}
Assume that we are given a set of sleep datasets $D=\{d_1,d_2,...,d_N\}$, where $N$ is the total number of datasets.
Our goal is to computationally quantify the impact of various data characteristics and the transferability (i.e., usefulness) between these datasets by utilizing the performance metrics obtained from different transfer configurations (Section~\ref{sec:tf_setting}).
Such configurations are designed to study how the model performance changes when we introduce different data characteristics between the source and target of transfer (RQ1) (Section~\ref{sec:method_qty_impact}).
This also allows us to determine whether transfer learning can improve the performance across all transfer configurations and which sources of transfer are more useful to target datasets compared to others (RQ2) (Section~\ref{sec:method_qty_trans}).

\subsection{Datasets}
In this study, we extracted six single-channel EEG datasets from four sleep datasets: (1) Montreal Archive of Sleep Studies (MASS)~\cite{oreilly2014}, (2) Sleep-EDF~\cite{kemp2000,goldberger2000}, (3) ISRUC-Sleep~\cite{khalighi2016}, and (4) Sleep Heart Health Study (SHHS) Visit~1~\cite{zhang2018} to evaluate our method.
Table~\ref{tab:datasets} summarizes the characteristics and the available EEG channels of the datasets.
These datasets are publicly available and commonly used to evaluate sleep stage scoring models.
They were also collected in different environments and annotated with different sleep manuals, which can be used to simulate different scenarios to evaluate the effectiveness of our method.

\begin{table*}[!t]
\centering
\captionsetup{width=\linewidth}
\caption{Characteristics of public sleep datasets used in this study, including the single-channel EEGs used as the source and target for transfer learning, the recording environments, and the distribution of sleep stages of each dataset.
Note that F3/F4 and C3/C4 indicate that F3 and C3 were used as the source dataset only when the target datasets were F4 and C4 from the same dataset. This is to avoid using the exact same dataset for the source and target and to simulate sensor placement variation in practice.}
\label{tab:datasets}
\renewcommand{\arraystretch}{1.15}
\resizebox{1.0\linewidth}{!}{%
\begin{tabular}{|l|c|c|c|c|c|c|c|c|c|c|c|c|c|}
\hline
\multirow{2}{*}{\textbf{Dataset}} &
  \multicolumn{2}{c|}{\textbf{EEG Channels}} &
  \multicolumn{1}{l|}{\multirow{2}{*}{\textbf{Manual}}} &
  \multicolumn{1}{l|}{\multirow{2}{*}{\textbf{Conditions}}} &
  \multicolumn{1}{l|}{\multirow{2}{*}{\textbf{Subjects}}} &
  \multicolumn{1}{l|}{\multirow{2}{*}{\textbf{Sampling Rate}}} &
  \multicolumn{1}{l|}{\multirow{2}{*}{\textbf{Epoch Length}}} &
  \multicolumn{6}{c|}{\textbf{Epochs}} \\ \cline{2-3} \cline{9-14} 
 &
  \textbf{Source} &
  \textbf{Target} &
  \multicolumn{1}{l|}{} &
  \multicolumn{1}{l|}{} &
  \multicolumn{1}{l|}{} &
  \multicolumn{1}{l|}{} &
  \multicolumn{1}{l|}{} &
  \textbf{W} &
  \textbf{N1} &
  \textbf{N2} &
  \textbf{N3} &
  \textbf{REM} &
  \textbf{Total} \\ \hline
MASS-SS1 & F3/F4,C3/C4,Pz,O2 & F4,C4 & AASM & Healthy & 53 & 256 & 30 & 12242 & 7112 & 22167 & 3407 & 6365 & 51293 \\ \hline
MASS-SS3 & F3/F4,C3/C4,Pz,O2 & F4,C4 & AASM & Healthy & 62 & 256 & 30 & 6442 & 4839 & 29802 & 7653 & 10581 & 59317 \\ \hline
Sleep-EDF-SC & Fpz-Cz,Pz-Oz & Fpz-Cz & R\&K & Healthy & 20 & 100 & 30 & 10197 & 2804 & 17799 & 5703 & 7717 & 44220 \\ \hline
ISRUC-SG1 & F3/F4,C3/C4,O2 & F4,C4 & AASM & Sleep Apnea & 100 & 200 & 30 & 20098 & 11062 & 27511 & 17251 & 11265 & 87187 \\ \hline
SHHS1-Normal & C3/C4 & C4 & R\&K & Normal & 100 & 125 & 30 & 27324 & 2986 & 40080 & 16103 & 15637 & 102130 \\ \hline
SHHS1-OSA & C3/C4 & C4 & R\&K & Sleep Apnea & 100 & 125 & 30 & 30326 & 3805 & 40556 & 12993 & 13583 & 101263 \\ \hline
\end{tabular}%
}
\end{table*}

\subsubsection{MASS}
In MASS cohort 1, there are PSG recordings from 200 subjects aged 18-76 years (97 males and 103 females) from different sleep research laboratories.
These recordings are organized into five subsets, SS1-SS5, according to their research and acquisition protocols.
The PSG recordings were segmented into 30 s epochs (SS1 and SS3) or 20 s epochs (SS2, SS4, and SS5).
Such PSG epochs were manually labeled by experts according to the AASM manual~\cite{iber2007} (SS1 and SS3) or the R\&K manual~\cite{rechtschaffen1968} (SS2, SS4, and SS5).
We converted the recordings annotated with the R\&K manual (W, N1, N2, N3, N4, and REM) to be the same as the AASM manual (W, N1, N2, N3, and REM) by merging the N3 and N4 stages into a single stage N3.
This is to facilitate a comparison of the sleep stage scoring performance across datasets, as recommended by Imtiaz and Rodriguez-Villegas~\cite{imtiaz2014}.
Each PSG recording consists of EEG, EOG, EMG and ECG signals with varying channel counts and sampling rates across subsets.
This study uses EEG signals from different brain areas from the subsets with 30 s PSG epochs (i.e., MASS-SS1 and MASS-SS3).
All EEG signals in this dataset have a sampling rate of 256 Hz.

\subsubsection{Sleep-EDF}
The Sleep-EDF dataset contains 197 whole-night sleep PSG recordings from two studies: age effect in healthy subjects (SC) and temazepam effects on sleep (ST).
In this study, we used 153 PSG recordings from the SC study, collected from 78 subjects aged 25-101 years (37 males and 41 females).
The PSG recordings were segmented into 30 s epochs and manually annotated by experts according to the R\&K manual (W, N1, N2, N3, N4, and REM).
Similar to the MASS dataset, we also merged the N3 and N4 stages into a single stage N3 to use the AASM standard.
Each recording consists of two EEG channels: Fpz-Cz and Pz-Oz EEG, with a sampling rate of 100 Hz.
There were long awake (W) periods at the start and the end of each recording.
We only included 30 min of such periods immediately preceding and following the sleep periods, as our focus is on the sleep periods.
We also used version 1 of the Sleep-EDF dataset published in 2013 before the expansion, in which there are 39 PSG recordings from the SC study, collected from 20 subjects (i.e., Sleep-EDF-SC).

\subsubsection{ISRUC-Sleep}
This dataset provides all-night PSG recordings with a duration of approximately eight hours collected from human adults, including healthy subjects and subjects with sleep disorders under the influence of sleep medication.
There are 108 adult subjects with evidence of having sleep disorders, of which 100 of them (aged 51$\pm$16) had one recording session, and the other 8 (aged 46.87$\pm$18.7) had two recording sessions.
There are also ten healthy control subjects (aged 40$\pm$10) with one recording session to compare the recordings from the subjects with sleep disorders.
Each recording consists of signals from 19 channels, including EEG, EOG, EMG, and ECG.
All EEG, EOG, and chin EMG signals had a sampling rate of 200 Hz.
All recordings were visually scored by two different sleep experts in the Sleep Medicine Centre of the Hospital of Coimbra University (CHUC) according to the AASM standard (W, N1, N2, N3 and REM).
In this study, we used the EEG signals from 100 subjects with one recording session, named ISRUC-SG1, representing the majority of the subjects with sleep disorders. This was done to simulate scenarios when the source and target datasets for transfer have different subject conditions.

\subsubsection{SHHS1}
The SHHS Visit 1 dataset (SHHS1) is a sleep dataset from a multi-center cohort study named the Sleep Heart Health Study (SHHS)~\cite{zhang2018}.
The dataset was collected to determine whether sleep-related breathing, such as sleep apnea, is associated with an increased risk of coronary heart disease, stroke, all-cause mortality, and hypertension in the United States.
The AASM recommended apnea-hypopnea index (AHI) was calculated for each subject.
It can be used to classify subjects into four obstructive sleep apnea (OSA) severity classes: normal (AHI $<5$ events/h), mild (AHI $5$ to $<15$ events/h), moderate (AHI $15$ to $<30$ events/h) and severe (AHI $\geq30$ events/h)~\cite{budhiraja2019}.
The dataset contains all-night 30 s PSG recordings from 6,441 males and females older than 40 years.
Each recording contains only two EEG channels: C3 and C4, with sampling rates 125 Hz.
Each recording was segmented into 30 s epochs and was manually annotated by a single sleep expert according to the R\&K manual.
The R\&K annotations were converted into the AASM standard by merging the stages N3 and N4 into a single stage, N3.
To study the effect of the subject conditions on the model performance and transferability, we used a subset of 200 subjects randomly sampled from different OSA classes: 100 normal (whose AHI $<5$) and 100 OSA subjects (whose AHI $\geq5$).
As a result, we have two datasets, SHHS-1 Normal and SHHS-1 OSA, with the same recording channels and environments but different subject conditions.

For each dataset, we excluded movement artifacts at the beginning and the end of each recording labeled as MOVEMENT or UNKNOWN, as they did not belong to the five sleep stages.
It should also be emphasized that we did not apply any preprocessing techniques (e.g., filtering and mean normalization) to the original signals.
Table~\ref{tab:datasets} summarizes the number of subjects and the distribution of sleep stages of each dataset.

\subsection{Transfer Configurations}
\label{sec:tf_setting}
This study considers three data characteristics: recording channels, recording environments, and subject conditions.
These different characteristics are simulated by choosing the source and the target for transfer from available datasets (see Table~\ref{tab:datasets}) as follows:

\begin{itemize}
    \item \textbf{Different recording channels:} Two datasets are chosen such that the recording channels are from different brain areas (i.e., locations). For example, if the source is from the frontal area EEG-F3/F4, then the target will be from one of the other areas, such as EEG-C3/C4, Pz, or O2.
    \item \textbf{Different recording environments}: Two datasets are chosen from different sleep datasets. For example, if the source is the MASS-SS1 dataset, then the target will be from one of the others datasets, such as MASS-SS3, Sleep-EDF-SC, or ISRUC-SG1.
    \item \textbf{Different subject conditions:} Two datasets are chosen from sleep datasets containing different subject conditions.
    There are two conditions in this study: healthy (MASS-SS1, MASS-SS3, Sleep-EDF-SC, and SHHS1-Normal) and sleep apnea (ISRUC-SG1 and SHHS1-OSA).
    For example, if the source is the SHHS1-OSA dataset, then the target will be from one of the MASS-SS1, MASS-SS3, Sleep-EDF-SC, or SHHS1-Normal datasets.
\end{itemize}

The number of different data characteristics in each transfer configuration can vary from zero (i.e., no difference) to at most three (i.e., all differences).
When there is no difference between the source and target (i.e., the recording channels, environment, and subject conditions are similar), the source and target are chosen from the dataset that contains alternative EEG channels from the same brain area.
This depends on the channel availability of the sleep dataset.
For instance, when the source is the EEG-F4 from MASS-SS3, the target will be the EEG-F3 from MASS-SS3.
This is not possible when the source is the EEG-Fpz-Cz from Sleep-EDF-SC, as there is no alternative EEG from the same locations.
It was decided not to use the exact same dataset for both the source and target, as the sensor placement may differ in practical applications.
On the other extreme, when all of the considered data characteristics are different, the source and target datasets are from different brain areas and different datasets with different conditions.
For example, when the source is the EEG-C4 from SHHS1-OSA, the target will be the EEG-F4 from MASS-SS3.

\textbf{Sampling Procedure:} It is important to note that the source $S$ and the target $T$ datasets for transfer are sampled from $D$ (i.e., $S \subset D$ and $T \subset D$).
We only consider some possible transfer pairs, not an exhaustive list.
Several studies~\cite{tsinalis2016,supratak2017,sors2018} have shown that deep learning models based on single-channel EEG from the frontal and central lobes can score sleep stages better than models based on other head locations.
Thus, the target datasets are limited to the frontal (F4 and Fpz) and central (C4) lobe channels.

\begin{figure}[t!]
    \centering
    \begin{subfigure}[b]{0.49\textwidth}
        \centering
        \includegraphics[width=0.9\textwidth,trim={7cm 0.5cm 7cm 0.5cm},clip]{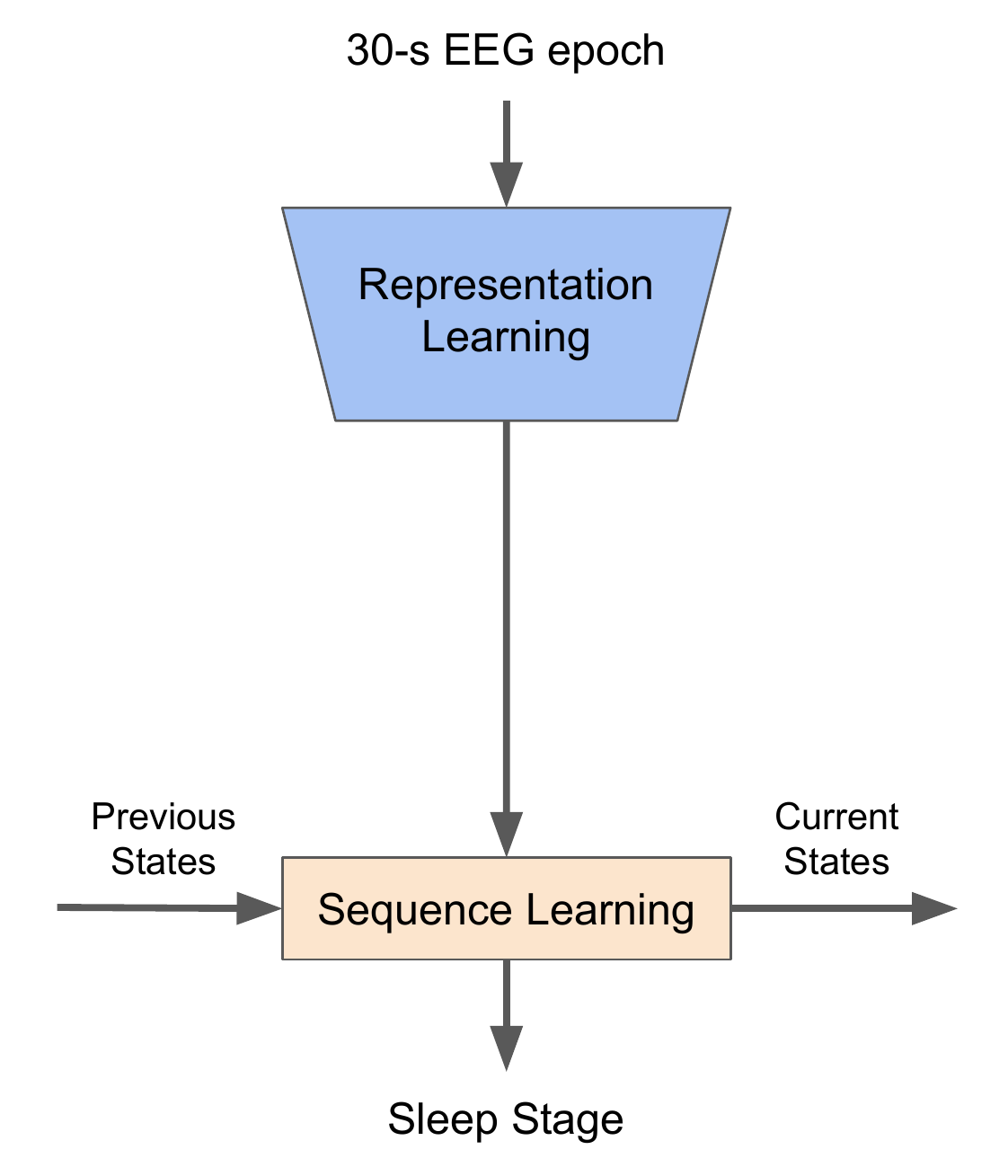}
        \caption{TinySleepNet.}
        \label{fig:tinysleepnet}
     \end{subfigure}
     \hfill
     \begin{subfigure}[b]{0.49\textwidth}
        \centering
        \includegraphics[width=0.9\textwidth,trim={7cm 0.5cm 7cm 0.5cm},clip]{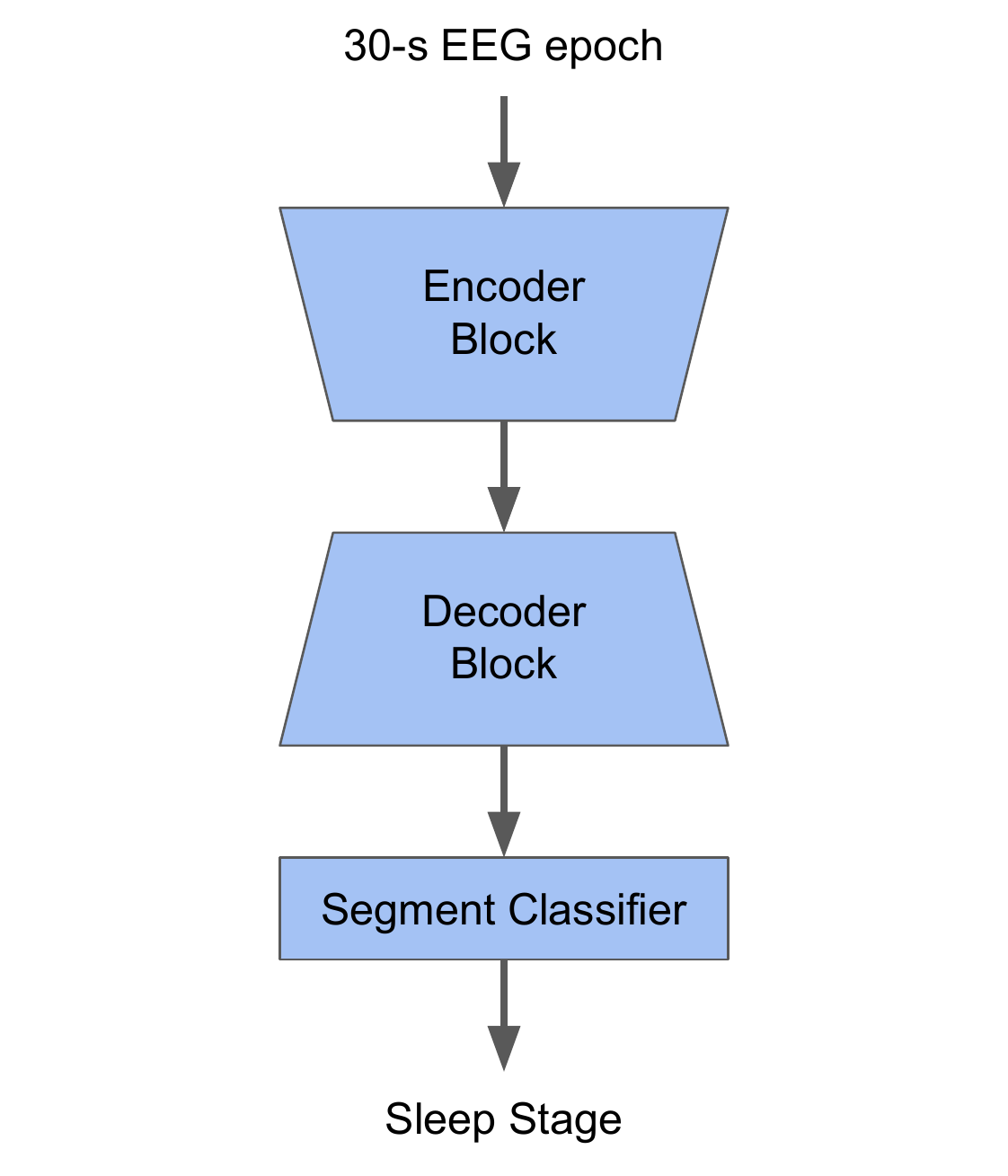}
        \caption{U-Time.}
        \label{fig:utime}
     \end{subfigure}
    \captionsetup{width=\linewidth}
    \caption{Overall architectures of TinySleepNet~\cite{supratak2020} and U-Time~\cite{perslev2019utime}. The blue and the orange colors are used to indicate the convolutional and the recurrent architecture, respectively. TinySleepNet utilizes both the convolutional and recurrent architectures to learn better representations and transition rules, while U-time utilizes the fully-convolutional encoder-decoder for dense classifications followed by an aggregation module for sleep stage scoring.}
    \label{fig:tinysleepnetutime}
\end{figure}

\subsection{Deep Learning Model} \label{sec:method_models}
Two deep learning models that can operate on raw EEG signals directly: TinySleepNet~\cite{supratak2020} and U-Time~\cite{perslev2019utime}, are used to evaluate the proposed method.
The former utilizes a convolutional and recurrent architectures to learn effective representations and transition rules.
The latter utilizes a fully-convolutional encoder-decoder for dense classifications and an aggregation module for sleep stage scoring.

\subsubsection{TinySleepNet}
TinySleepNet is a simple and efficient deep learning model~\cite{supratak2020} consisting of two main components: convolutional and recurrent modules (see Fig.~\ref{fig:tinysleepnet}).
The convolutional module learns useful representations of raw EEG signals, while the recurrent module learns transition rules among the sequence of representations for sleep stage scoring.
Due to signal and sequence augmentation during training, this model is robust to shifts along the time axis resulting from a difference in data collection protocols for segmenting long recordings into epochs for annotating sleep stages and is not prone to overfitting to sleep stage sequences.
It has also been shown to generalize well to sleep datasets with different characteristics (such as sampling rates, recording channels and environments, and sleep manuals).
The code is publicly available at https://github.com/akaraspt/tinysleepnet.

\subsubsection{U-Time}
U-Time is a deep learning model that employs a fully-convolutional encoder-decoder network~\cite{perslev2019utime} for dense segmentation of an input PSG epoch. In contrast to TinySleepNet, the U-Time model does not use any recurrent architecture to utilize the inter-epoch temporal information for prediction.
Instead, it trains an aggregation function to summarize the segmentation results into a sequence of sleep stages.
It has also been demonstrated to generalize well to different sleep datasets without any architecture and hyperparameter fine-tuning.
The code is publicly available at https://github.com/perslev/U-Time.

\subsection{Transfer Learning} \label{sec:method_transfer_learning}
Transfer learning is a technique that utilizes existing large (public) datasets to improve the model performance on a target dataset when the amount of annotated data is limited~\cite{phan2019chmismatch,zhuang2021}.
Typically, it consists of two main steps: pre-training and fine-tuning.
The pre-training step trains the model on the existing large dataset, resulting in an initial model that is fine-tuned in the next step.
The fine-tuning step trains the pre-trained model on the target dataset.
It has been shown that such a pre-trained model can provide a better starting point for fine-tuning the model on the target dataset, compared to random weight initialization.
We use this technique to help quantify the transferability of a model, as a model with better transferability should have a higher performance gain after the transfer compared to others.

\subsubsection{Pre-training} \label{sec:pretrain}
We trained a model from scratch in a supervised setting using the training set from each dataset $d_i$ to maximize the sleep stage scoring performance.
The model that performed best on the validation set from $d_i$ during the training was used as a pre-trained model $m_i$ for $d_i$.
This resulted in $N$ pre-trained models $M=\{m_1,m_2,...,m_N\}$, which were used for transfer learning.
The number of pre-trained models $N$ equals the number of source datasets, which will be discussed in Section~\ref{sec:experiment_setup}.

\subsubsection{Fine-tuning}
The pre-trained model $m_i$ was fine-tuned or transferred to the target dataset $d_j$ using its training set.
In particular, the pre-trained model $m_i$ (obtained from the pre-training step in Section~\ref{sec:pretrain}) was used as the starting point for training with the training set of the target dataset $d_j$.
There was no freezing layer during fine-tuning, therefore all model parameters (i.e., weights and biases) are adjusted.
The model that performed best on the validation set from $d_j$ during training was used as a fine-tuned or transferred model.
We denote the fine-tuned model from $d_i$ to $d_j$, as $m_{i \to j}$.
This fine-tuning step was applied for all feasible source-target pairs as specified in Table~\ref{tab:datasets}.
This resulted in a set of fine-tuned models for each target $d_j$, $\{m_{1 \to j},m_{2 \to j},...,m_{N \to j}\}$.

\subsubsection{Implementation}
The original code provided by the authors of the models (see Section~\ref{sec:method_models}) was used to implement both the pre-training and the fine-tuning steps without any modifications, unless additional configurations were required to run on new datasets.
The only difference between the pre-training and fine-tuning was the initial weights, with the former using random weights and the latter using the pre-trained weights.
The default hyperparameters for training (e.g., learning rate, weight decay, and training epochs) were used.
No babysitting mechanism was employed during the training to obtain the best results, but rather, the best parameters were chosen based on the validation set.
\subsection{Quantifying Impact of Different Data Characteristics}
\label{sec:method_qty_impact}
To better understand which data characteristics between the source datasets and the target dataset affect the performance the most (RQ1), we propose using a relative performance difference to rank the data characteristics that affect the performance the most.
The absolute performance differences cannot be used to answer this question as they are in different ranges across sleep datasets, as are their differences.

In particular, the impact of each different data characteristic can be quantified by computing an average of relative performance differences between the pre-trained models, $M$, from all pairs of datasets with such different characteristics.
Formally, given a pair of datasets: $d_i$ and $d_j$, the relative performance difference of these two, $r_{i,j}$, is the percentage:
\begin{equation} \label{eq:impact_factor}
    r_{i,j} = \left( \frac{p(m_j, d_j)}{p(m_i, d_j)} - 1 \right) \times 100\%,
\end{equation}
where $p(m, d)$ is the performance when applying the model $m$ on the test set from $d$.
The quantity $r_{i,j}$ is positive when $p(m_j,d_j) > p(m_i,d_j)$, indicating a degradation in performance when the model trained on one dataset (i.e., $d_i$) is applied to another dataset (i.e., $d_j$).
The quantity $r_{i,j}$ is negative when $p(m_j,d_j) < p(m_i,d_j)$, indicating a performance gain.
The higher the magnitude of $r_{i,j}$, the more the impact of such characteristics on model performance.
We can estimate the impact of using the pre-trained models from the datasets that have, for example, different recording channels from the target datasets using an average of $r_{i,j}$ over all pairs of $d_i$ and $d_j$, where only the recording channels differ (see Table~\ref{tab:perf_diff_tf_configs}).

\subsection{Quantifying Transferability}
\label{sec:method_qty_trans}
To determine which dataset should be used as the source of transfer to maximize model performance on the target dataset (RQ2), we need a transferability matrix, $W$, where each element $w_{i,j}$ represents the transferability or the usefulness of a source $d_j$ to a target $d_i$.
Such a matrix enables us to determine which source datasets are most useful for a target dataset.

We adapted an ordinal approach, called the analytic hierarchy process (AHP)~\cite{saaty1987}, commonly used to study the relations of multiple criteria for decision making, to normalize the performance differences from multiple pairwise comparisons into ratio scales, which can be used to rank datasets based on transferability.

In particular, we first constructed a pairwise comparison matrix between feasible source datasets for transferring to a target dataset $d_t$, named $H_t$.
Each element at $(i,j)$ is the percentage performance difference relative to the target $d_t$ for which source $d_i$ transferred to $d_t$ did better than source $d_j$ did:
\begin{equation} \label{eq:transferability}
    h_t^{(i,j)} = \frac{p(m_{i \to t}, d_t) - p(m_{j \to t}, d_t)}{p(m_t, d_t)} \times 100\%.
\end{equation}

\begin{figure*}[!t]
  \begin{subfigure}[b]{0.49\linewidth}
    \centerline{\includegraphics[width=\linewidth]{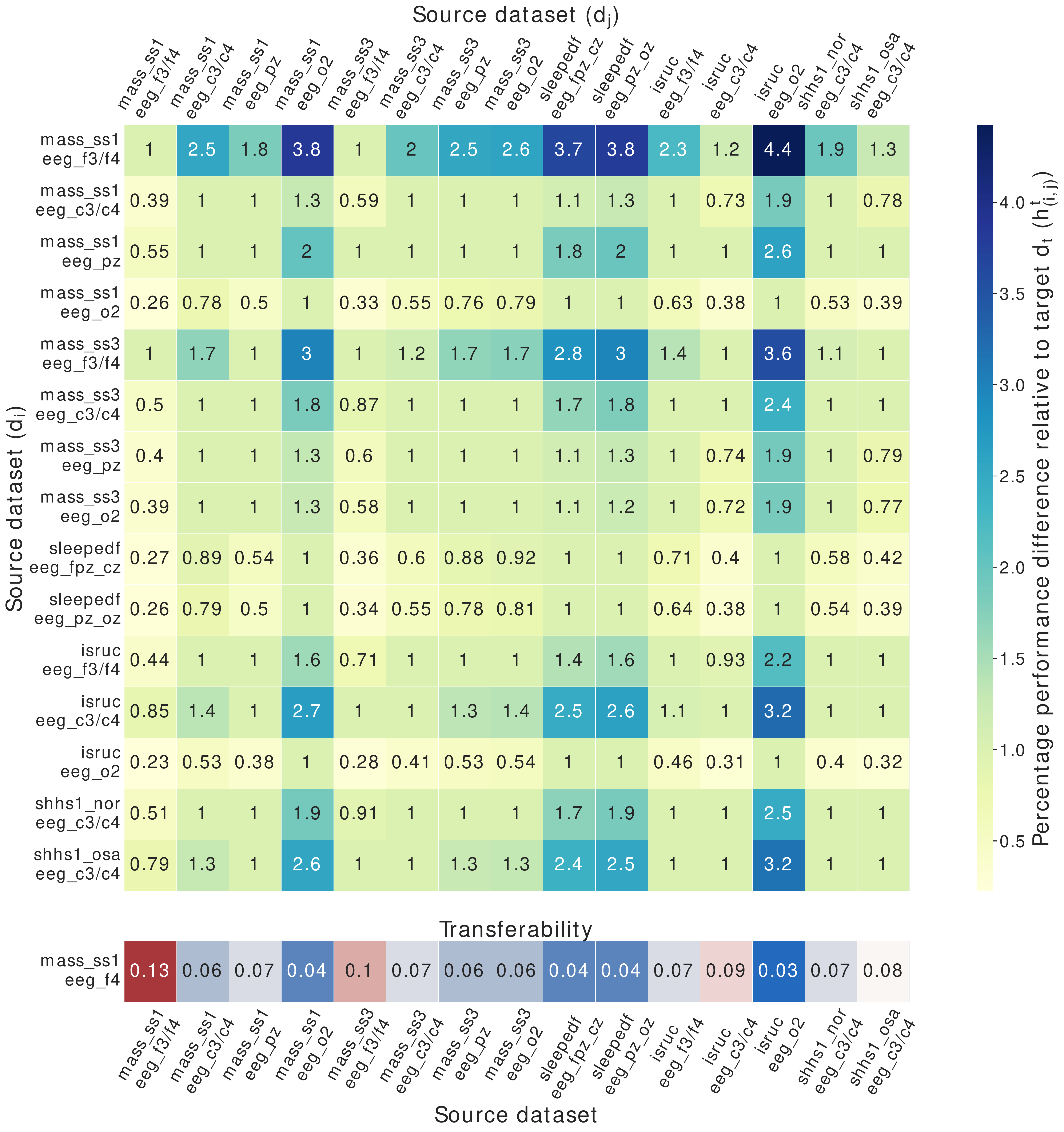}}
    \caption{$H_t$ of the EEG-F4 from MASS-SS1 from TinySleepNet.}
    \label{fig:result_pairwise_comp_tinysleepnet}
  \end{subfigure}
  \hfill 
  \begin{subfigure}[b]{0.49\linewidth}
    \centerline{\includegraphics[width=\linewidth]{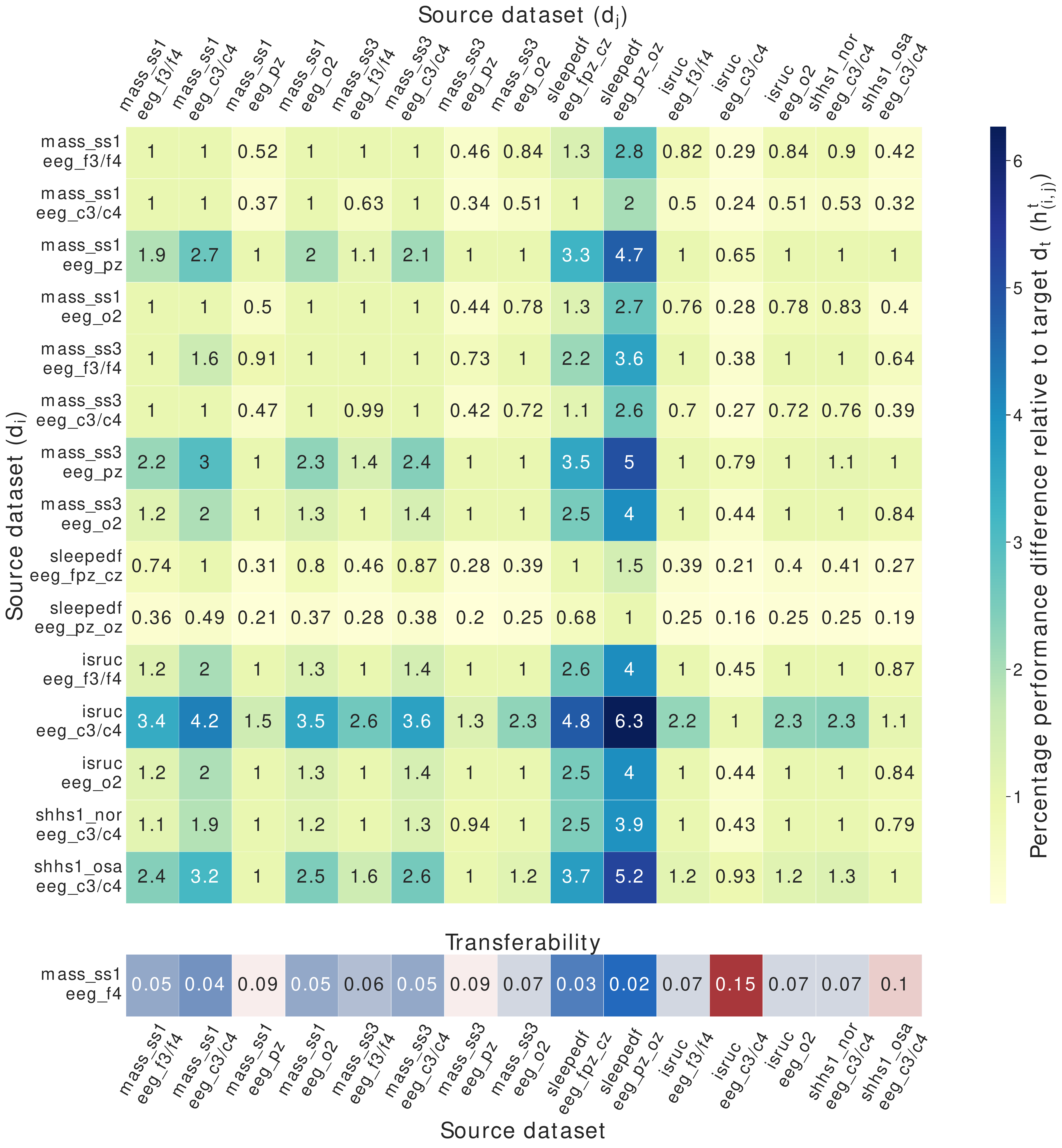}}
    \caption{$H_t$ of the EEG-F4 from MASS-SS1 from U-Time.}
    \label{fig:result_pairwise_comp_utime}
  \end{subfigure}
  \captionsetup{width=\linewidth}
  \caption{Examples of pairwise comparison matrices ($H_t$) of the EEG-F4 from MASS-SS1 from TinySleepNet (a) and the EEG-F4 from MASS-SS1 from U-Time (b) datasets.
  When $h_t^{(i,j)}>1$, it means that the model transferred from $d_i$ score sleep stages on dataset $d_t$ is better than $d_j$ (i.e., darker color).
  When $h_t^{(i,j)}<1$, it means that the model transferred from $d_j$ score sleep stage on dataset $d_t$ is better than $d_i$ (i.e., lighter color).
  When $h_t^{(i,j)}=1$, it means that both the models transferred from $d_i$ and $d_j$ perform equally well on dataset $d_t$ (i.e., green).
  The two vectors at the bottom are the corresponding eigenvectors representing the transferability of each source dataset.
  The sources of transfer with high and low degrees of transferability are highlighted in red and blue, respectively.}
  \label{fig:result_pairwise_comp}
\end{figure*}

After we calculated $h_t^{(i,j)}$ for all possible pairs, we further normalized $H_t$ as follows:
\begin{equation} \label{eq:tf_normalization}
    h_t^{(i,j)} = 
\begin{cases}
    1,& \text{if } i = j\\
    h_t^{(i,j)},& \text{if } h_t^{(i,j)} > \alpha\\
    \frac{1}{ \left| h_t^{(j,i)} \right| },& \text{if } h_t^{(i,j)} < -\alpha\\
    1,& \text{otherwise}
\end{cases}
\end{equation}
where $\alpha$ is the minimum percentage difference that will be used to judge whether one source is better than another.
In this study, we chose $\alpha=1$ in the normalization process as it was less sensitive to noise and helped improve the numerical stability when the performance difference was very small.
Such normalization produces a matrix that has a similar interpretation to the ones introduced in the original AHP process, which represents the importance of each criterion/objective before extracting an eigenvector for each dataset.
The normalized $H_t$ matrix can be interpreted as follows:
\begin{itemize}
    \item When $h_t^{(i,j)}>1$, it means that the model transferred from $d_i$ scores sleep stages on dataset $d_t$ is better than $d_j$ does.
    \item When $h_t^{(i,j)}<1$, it means that the model transferred from $d_j$ scores sleep stage on dataset $d_t$ better than $d_i$ does.
    \item When $h_t^{(i,j)}=1$, it means that both the models transferred from $d_i$ and $d_j$ perform equally well on dataset $d_t$.
\end{itemize}

Fig.~\ref{fig:result_pairwise_comp} illustrates two examples of $H_t$ calculated for TinySleepNet and U-Time when the target dataset was the EEG-F4 from MASS-SS1.
For the TinySleepNet model (Fig.~\ref{fig:result_pairwise_comp_tinysleepnet}), the EEG-F3/F4 from MASS-SS1 was a better source of transfer compared to most of the others, as most of the $h^t_{(i,j)}$ elements in the first row are greater than 1.
The EEG-O2 from ISRUC-SG1, on the other hand, was the least useful one, as most of the $h^t_{(i,j)}$ elements in its row are less than 1.
The most and the least useful transfer sources for the EEG-F4 from MASS-SS1 were the EEG-F3/F4 from MASS-SS1 and the EEG-O2 from ISRUC-SG1.
For the U-Time model (Fig.~\ref{fig:result_pairwise_comp_utime}), the most and the least useful transfer sources for the EEG-F4 from MASS-SS1 were the EEG-C3/C4 from ISRUC-SG1 and the EEG-Pz-Oz from Sleep-EDF-SC, respectively. 

Once we have $H_t$ for all target datasets, the transferability of each source dataset to the target $d_t$ is determined by calculating the principal eigenvector of $H_t$ in an approximate manner to simplify the calculation process~\cite{vargas2010}.
The $i$-th component of the approximated eigenvector indicates the transferability of $d_i$ relative to $d_t$.
Two examples of the eigenvectors are shown at the bottom of Fig.~\ref{fig:result_pairwise_comp}.
As each eigenvector represents the transferability of different sources to a target, we can stack multiple principal eigenvectors of $H_t$ for all $d_t \in T$ to get a transferability matrix $W$ (see Fig.~\ref{fig:result_transferability}).

\section{Results}

\subsection{Experimental Setup} \label{sec:experiment_setup}
We evaluated the proposed method on TinySleepNet and U-Time models and compared the impact of data characteristics on transferability.
With six sleep datasets, there were 23 pre-training and 134 fine-tuning runs for each of TinySleepNet and U-Time.
The number of pre-trained models equaled the number of EEG channels from all source datasets (see Table~\ref{tab:datasets}).
It should be noted that EEG-F3, F4, C3, and C4 were trained separately.
The number of fine-tuned models was obtained from the number of sources (i.e., $23-8$) multiplied by the number of targets (i.e., $9$) and reduced by $1$ (i.e., $((23-8)*9)-1=134$).
The number of possible transfer sources was reduced by $8$ since EEG-F3 and C3 were only used when the target dataset was EEG-F4 and C4 from the same dataset, respectively.
The reduction by $1$ was due to no alternative EEG channel from the frontal area in Sleep-EDF-SC.
The training was repeated three times, and the averages of the repeated training in each configuration were used.
The total training time was approximately 1300 GPU hours (or 8 weeks) on 4 NVIDIA GTX-1080ti GPUs.

\textbf{Train/Test Split.} We randomly split each dataset into \textit{subject-independent} training and test sets (80/20).
Such an independent split ensured that there were no PSG epochs from the same subjects in both sets.
The training set was further split into training and validation sets (90/10), and the validation set was used to select the best model during the training (as mentioned in Section~\ref{sec:method_transfer_learning}).
Note that the test sets from all datasets were the same for all transfer configurations to ensure fair performance comparisons.

\textbf{Training Models.}
Both models were trained using the same architectures and hyperparameters as in~\cite{supratak2020,perslev2019utime}.
The input size of the TinySleepNet model depended on the sampling rate of EEG signals from each dataset.
During transfer learning, if the input size of the target dataset was not the same (e.g., Sleep-EDF-SC and MASS-SS1), the EEG signals of the target dataset were resampled using the Fourier method to be the same size as the model's input.
With the Fourier method, the signals were transformed into the frequency domain to up/down-sample before being transformed back to the time domain.
The input size of the U-Time model was fixed to the sampling rate of 128 Hz, and all input EEG signals were resampled using polyphase filtering with automatically derived FIR filters.

\textbf{Performance Metrics.} 
We evaluated the sleep stage scoring performance using the overall accuracy (ACC) and the macro-average F1-score (MF1), which has been widely used for performance comparison.
The MF1 can be computed as follows:
\begin{equation}
    \text{MF1} = \frac{\sum_{c=1}^{C} \text{F1}_c}{C} ,
\end{equation}
where F1$_c$ is a per-class F1-score of class $c$ and $C$ is the number of sleep stages (which is 5).
The performance metric used in Eq.\ref{eq:impact_factor} and Eq.\ref{eq:transferability} is MF1, which is more appropriate than ACC for datasets with imbalanced classes.

\textbf{Evaluation Settings.} We evaluated the model performance in three settings as follows:
\begin{itemize}
    \item \textit{From Scratch (FS)}. Performance from applying a pre-trained model from a target dataset ($m_t$) on the test set of the target dataset ($p(m_t,d_t)$).
    \item \textit{Direct Transfer (DT)}. Performance from applying a pre-trained model from a source dataset ($m_s$) on the test set of the target dataset ($p(m_s,d_t)$), where $s \neq t$.
    \item \textit{Fine-tuned Transfer (FT)}. Performance from applying a fine-tuned model on a target dataset ($m_{s \to t}$) on the test set of the target dataset ($p(m_{s \to t},d_t)$), where $s \neq t$.
\end{itemize}

We quantified the impact of various data characteristics on the transferability between source and target datasets using FS versus DT, and FS versus FT, respectively.

\subsection{Sleep Stage Scoring Performance}
\begin{figure*}[!t]
  \centering
  \begin{subfigure}[b]{1.0\textwidth}
     \centering
     \includegraphics[width=\textwidth]{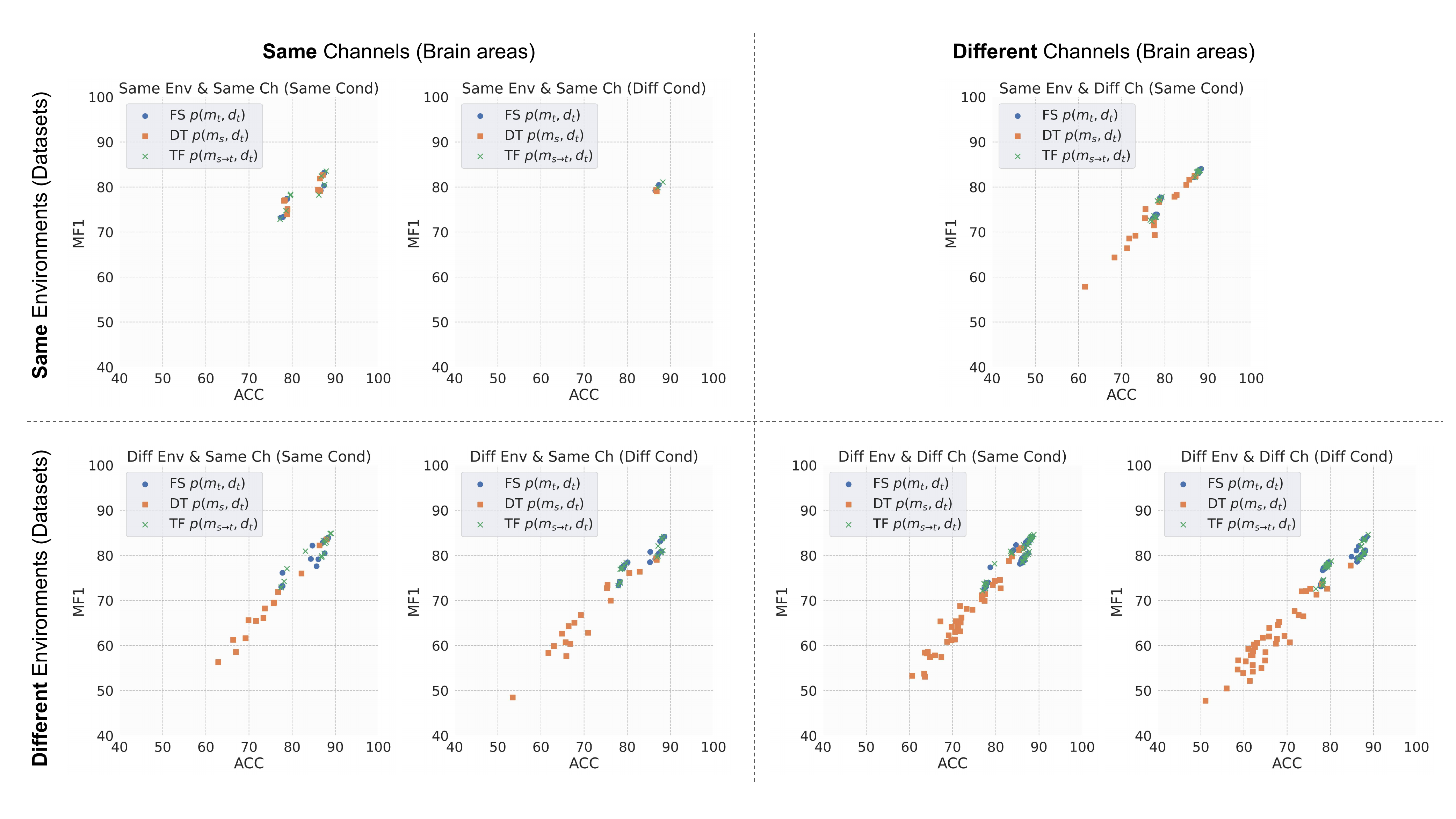}
     \caption{TinySleepNet.}
     \label{fig:score_performance_tinysleepnet}
  \end{subfigure}
  \vfill
  \begin{subfigure}[b]{1.0\textwidth}
     \centering
     \includegraphics[width=\textwidth]{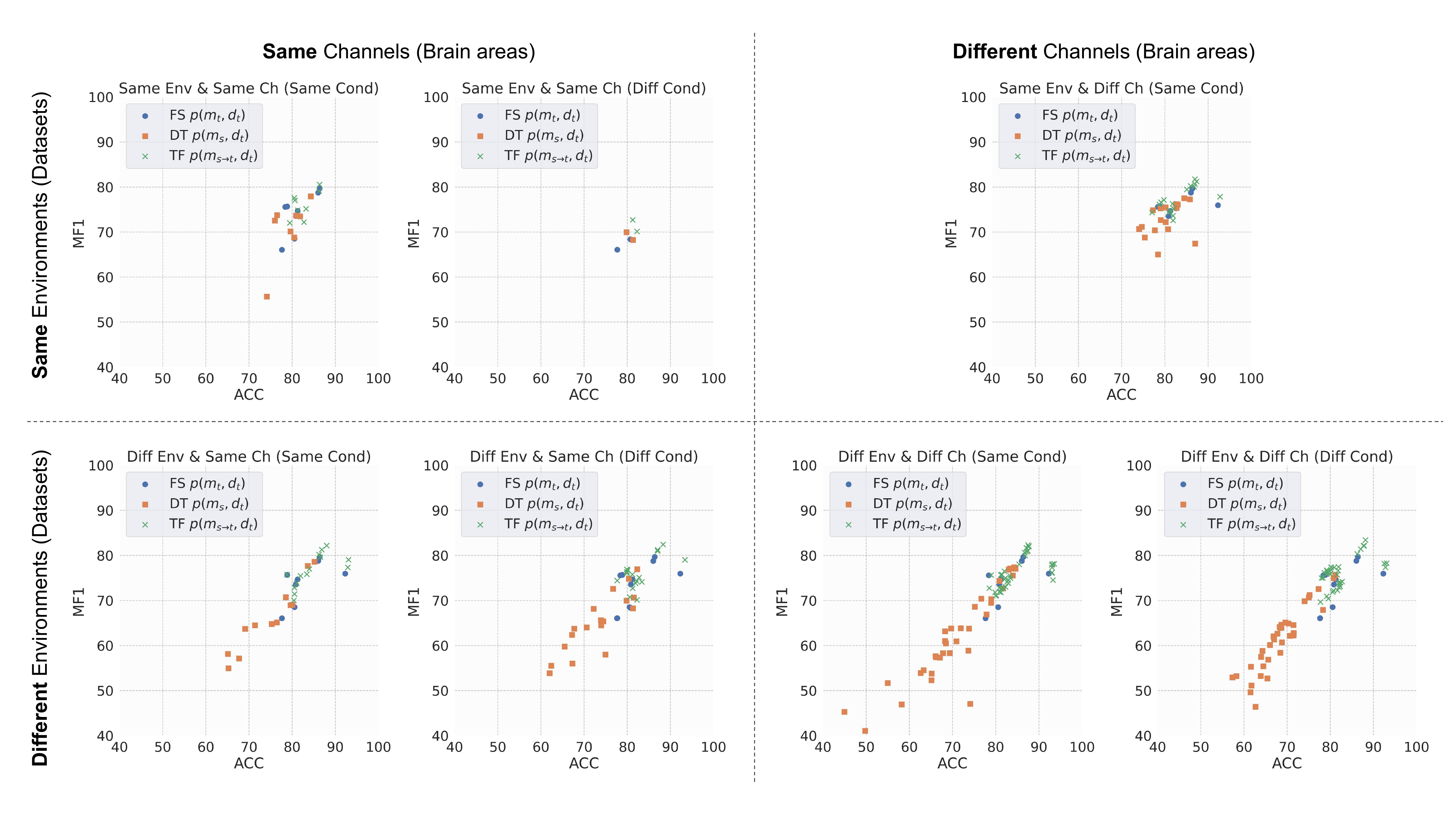}
     \caption{U-Time.}
     \label{fig:score_performance_utime}
  \end{subfigure}
  \captionsetup{width=\linewidth}
  \caption{Distribution of the ACC-MF1 performance from three evaluation settings: From Scratch (FS, blue dot), Direct Transfer (DT, orange square) and Fine-tuned Transfer (FT, green cross) in different transfer configurations for TinySleepNet (a) and U-Time (b).
  }
  \label{fig:score_performance}
\end{figure*}

Table~\ref{tab:perf_diff_tf_configs} shows the average of ACC and MF1 on FS, DT, and FT from different configurations for the TinySleepNet and U-Time models.
The distribution of each transfer configuration is illustrated in Fig.~\ref{fig:score_performance}.
The number of data points differed across configurations, depending on the possible source-target pairs for transfer in each configuration.

The ACC-MF1 performances from FS and DT were first compared using a simple performance difference.
When the recording channels and environments were the same, the ACC-MF1 performance of FS and DT were similar for TinySleepNet regardless of the subject conditions.
Unlike TinySleepNet, the DT performance of U-Time decreased less when the subject conditions were the same than when they were different.
When the recording environments were the same but with different channels, the DT performance decreased for both models even though the subject conditions were the same.
When the recording environments were different, a more significant ACC-MF1 performance gap between FS and DT was observed in both models, regardless of the channels and subject conditions.
For TinySleepNet, the DT performance decreased further when one more differences in either the recording channels or the subject conditions were introduced.
When all considered characteristics were different, the DT performance of TinySleepNet dropped the most.
For U-Time, one more difference in recording channels did not degrade the DT performance much when the subject conditions were the same.
Surprisingly, when the subject conditions were the same, the performance gap between FS and DT was more significant than when they were different.
The DT performance of U-Time dropped the most when the recording channels and environments were different, but the subject conditions were the same.
To identify the causes of the surprising results from U-Time performance, further investigation and babysitting the U-Time training are required. However, they are beyond the scope of this study, and will be investigated in the future.

The ACC-MF1 performances from FS, DT, and FT were further compared to investigate transfer learning results.
We found that fine-tuning the pre-trained model on the target dataset (FT) consistently achieved better performance than DT for both models, regardless of the different characteristics between the source and target.
Furthermore, most results showed that the FT performance was higher than FS, suggesting that transfer learning can help mitigate the performance drop across all configurations. 
Both TinySleepNet and U-Time were able to adapt to the target dataset.

\begin{table}[!t]
\centering
\captionsetup{width=\linewidth}
\caption{Average of overall accuracy (ACC) and macro-average F1-score (MF1) from different transfer configurations for TinySleepNet and U-Time models. From Scratch (FS) represents $p(m_t,d_t)$, Direct Transfer (DT) represents $p(m_s,d_t)$, and Fine-tuned Transfer (FT) represents $p(m_{s \to t},d_t)$. The numbers in bold indicate the best performance among FS, DT and FT for each model.}
\label{tab:perf_diff_tf_configs}
\renewcommand{\arraystretch}{1.15}
\resizebox{\columnwidth}{!}{%
\begin{tabular}{llll|ccccccc|ccccccc|}
\cline{5-18}
 &
   &
   &
   &
  \multicolumn{7}{c|}{TinySleepNet} &
  \multicolumn{7}{c|}{U-Time} \\ \cline{3-18} 
 &
  \multicolumn{1}{l|}{} &
  \multicolumn{1}{l|}{\multirow{2}{*}{Channels}} &
  \multirow{2}{*}{Conditions} &
  \multicolumn{2}{c|}{FS} &
  \multicolumn{2}{c|}{DT} &
  \multicolumn{2}{c|}{FT} &
  \multirow{2}{*}{$r_{i,j}$} &
  \multicolumn{2}{c|}{FS} &
  \multicolumn{2}{c|}{DT} &
  \multicolumn{2}{c|}{FT} &
  \multirow{2}{*}{$r_{i,j}$} \\ \cline{5-10} \cline{12-17}
 &
  \multicolumn{1}{l|}{} &
  \multicolumn{1}{l|}{} &
   &
  \multicolumn{1}{c|}{ACC} &
  \multicolumn{1}{c|}{MF1} &
  \multicolumn{1}{c|}{ACC} &
  \multicolumn{1}{c|}{MF1} &
  \multicolumn{1}{c|}{ACC} &
  \multicolumn{1}{c|}{MF1} &
   &
  \multicolumn{1}{c|}{ACC} &
  \multicolumn{1}{c|}{MF1} &
  \multicolumn{1}{c|}{ACC} &
  \multicolumn{1}{c|}{MF1} &
  \multicolumn{1}{c|}{ACC} &
  \multicolumn{1}{c|}{MF1} &
   \\ \hline
\multicolumn{1}{|l|}{\multirow{8}{*}{\rotatebox[origin=c]{90}{Environment}}} &
  \multicolumn{1}{l|}{\multirow{4}{*}{\rotatebox[origin=c]{90}{Same}}} &
  \multicolumn{1}{l|}{Same} &
  Same &
  \multicolumn{1}{c|}{82.7} &
  \multicolumn{1}{c|}{78.4} &
  \multicolumn{1}{r|}{82.5} &
  \multicolumn{1}{c|}{78.3} &
  \multicolumn{1}{c|}{\textbf{82.9}} &
  \multicolumn{1}{c|}{\textbf{78.6}} &
  1.0 &
  \multicolumn{1}{c|}{81.2} &
  \multicolumn{1}{c|}{74.0} &
  \multicolumn{1}{r|}{79.2} &
  \multicolumn{1}{c|}{70.7} &
  \multicolumn{1}{c|}{\textbf{82.6}} &
  \multicolumn{1}{c|}{\textbf{76.1}} &
  7.4 \\ \cline{3-18} 
\multicolumn{1}{|l|}{} &
  \multicolumn{1}{l|}{} &
  \multicolumn{1}{l|}{Same} &
  Different &
  \multicolumn{1}{c|}{86.9} &
  \multicolumn{1}{r|}{79.8} &
  \multicolumn{1}{c|}{86.8} &
  \multicolumn{1}{c|}{79.2} &
  \multicolumn{1}{c|}{\textbf{87.7}} &
  \multicolumn{1}{c|}{\textbf{80.5}} &
  1.0 &
  \multicolumn{1}{c|}{79.2} &
  \multicolumn{1}{r|}{67.2} &
  \multicolumn{1}{c|}{80.6} &
  \multicolumn{1}{c|}{69.1} &
  \multicolumn{1}{c|}{\textbf{81.8}} &
  \multicolumn{1}{c|}{\textbf{71.4}} &
  2.7 \\ \cline{3-18} 
\multicolumn{1}{|l|}{} &
  \multicolumn{1}{l|}{} &
  \multicolumn{1}{l|}{Different} &
  Same &
  \multicolumn{1}{c|}{\textbf{82.1}} &
  \multicolumn{1}{c|}{\textbf{78.4}} &
  \multicolumn{1}{c|}{77.5} &
  \multicolumn{1}{c|}{73.4} &
  \multicolumn{1}{c|}{82.0} &
  \multicolumn{1}{c|}{78.3} &
  7.4 &
  \multicolumn{1}{c|}{83.0} &
  \multicolumn{1}{c|}{76.4} &
  \multicolumn{1}{c|}{80.1} &
  \multicolumn{1}{c|}{72.8} &
  \multicolumn{1}{c|}{\textbf{83.3}} &
  \multicolumn{1}{c|}{\textbf{77.2}} &
  5.3 \\ \cline{3-18} 
\multicolumn{1}{|l|}{} &
  \multicolumn{1}{l|}{} &
  \multicolumn{1}{l|}{Different} &
  Different &
  \multicolumn{1}{c|}{-} &
  \multicolumn{1}{c|}{-} &
  \multicolumn{1}{c|}{-} &
  \multicolumn{1}{c|}{-} &
  \multicolumn{1}{c|}{-} &
  \multicolumn{1}{c|}{-} &
  - &
  \multicolumn{1}{c|}{-} &
  \multicolumn{1}{c|}{-} &
  \multicolumn{1}{c|}{-} &
  \multicolumn{1}{c|}{-} &
  \multicolumn{1}{c|}{-} &
  \multicolumn{1}{c|}{-} &
  - \\ \cline{2-18} 
\multicolumn{1}{|l|}{} &
  \multicolumn{1}{l|}{\multirow{4}{*}{\rotatebox[origin=c]{90}{Different}}} &
  \multicolumn{1}{l|}{Same} &
  Same &
  \multicolumn{1}{c|}{84.1} &
  \multicolumn{1}{c|}{79.4} &
  \multicolumn{1}{c|}{74.2} &
  \multicolumn{1}{c|}{68.3} &
  \multicolumn{1}{c|}{\textbf{84.6}} &
  \multicolumn{1}{c|}{\textbf{80.0}} &
  17.6 &
  \multicolumn{1}{c|}{83.7} &
  \multicolumn{1}{c|}{74.6} &
  \multicolumn{1}{c|}{69.5} &
  \multicolumn{1}{c|}{61.3} &
  \multicolumn{1}{c|}{\textbf{84.6}} &
  \multicolumn{1}{c|}{\textbf{76.5}} &
  30.2 \\ \cline{3-18} 
\multicolumn{1}{|l|}{} &
  \multicolumn{1}{l|}{} &
  \multicolumn{1}{l|}{Same} &
  Different &
  \multicolumn{1}{c|}{83.0} &
  \multicolumn{1}{c|}{78.8} &
  \multicolumn{1}{c|}{71.1} &
  \multicolumn{1}{c|}{66.3} &
  \multicolumn{1}{c|}{\textbf{83.2}} &
  \multicolumn{1}{c|}{\textbf{79.0}} &
  20.5 &
  \multicolumn{1}{c|}{81.1} &
  \multicolumn{1}{c|}{73.5} &
  \multicolumn{1}{c|}{73.0} &
  \multicolumn{1}{c|}{65.0} &
  \multicolumn{1}{c|}{\textbf{82.7}} &
  \multicolumn{1}{c|}{\textbf{76.0}} &
  14.6 \\ \cline{3-18} 
\multicolumn{1}{|l|}{} &
  \multicolumn{1}{l|}{} &
  \multicolumn{1}{l|}{Different} &
  Same &
  \multicolumn{1}{c|}{84.9} &
  \multicolumn{1}{c|}{80.0} &
  \multicolumn{1}{c|}{72.5} &
  \multicolumn{1}{c|}{66.0} &
  \multicolumn{1}{c|}{\textbf{85.1}} &
  \multicolumn{1}{c|}{\textbf{80.1}} &
  22.5 &
  \multicolumn{1}{c|}{84.1} &
  \multicolumn{1}{c|}{74.2} &
  \multicolumn{1}{c|}{67.6} &
  \multicolumn{1}{c|}{59.3} &
  \multicolumn{1}{c|}{\textbf{85.0}} &
  \multicolumn{1}{c|}{\textbf{76.2}} &
  31.6 \\ \cline{3-18} 
\multicolumn{1}{|l|}{} &
  \multicolumn{1}{l|}{} &
  \multicolumn{1}{l|}{Different} &
  Different &
  \multicolumn{1}{c|}{82.7} &
  \multicolumn{1}{c|}{78.8} &
  \multicolumn{1}{c|}{66.4} &
  \multicolumn{1}{c|}{61.6} &
  \multicolumn{1}{c|}{\textbf{83.0}} &
  \multicolumn{1}{c|}{\textbf{79.0}} &
  29.4 &
  \multicolumn{1}{c|}{80.9} &
  \multicolumn{1}{c|}{73.6} &
  \multicolumn{1}{c|}{67.1} &
  \multicolumn{1}{c|}{60.1} &
  \multicolumn{1}{c|}{\textbf{82.2}} &
  \multicolumn{1}{c|}{\textbf{76.0}} &
  25.2 \\ \hline
\end{tabular}%
}
\end{table}

It should be noted that we aim not to maximize the sleep stage scoring performance but rather to understand how different characteristics between the source and the target datasets affect the performance and to quantify the transferability or the usefulness of source datasets to target datasets.
Thus, the performance reported in this study should not be compared with the other studies.
In addition, there were no results for one configuration (i.e., same environment, different channels, and conditions) due to the lack of datasets mentioned in Section~\ref{sec:tf_setting}.
The next section will demonstrate quantifying the impact of different data characteristics, which is a more reliable approach than a simple ACC-MF1 performance difference.

\subsection{Impact of Different Data Characteristics}
To compare which data characteristic impacted the sleep stage scoring performance the most (RQ1), the relative performance differences ($r_{i,j}$ in Eq.~\ref{eq:impact_factor}) using the FS and DT performance were computed for TinySleepNet and U-Time.
Each model's $r_{i,j}$ column summarizes the averages of $r_{i,j}$ from different transfer configurations.

For TinySleepNet, when the channels and environments were the same, the $r_{i,j}$ were close to 0 regardless of the subject conditions (1.0\% vs. 1.0\%).
The $r_{i,j}$ increased to 7.4\% when the channels were different, even though the environments and the conditions were similar.
When the recording environment differed, the $r_{i,j}$ jumped to more than 17\% in all configurations.
In particular, the $r_{i,j}$ was 17.6\% when the channels and the conditions were similar and increased to 20.5\% and 22.5\% when the conditions and the channels were different.
This conformed with the previous results when the environments were the same: the recording channels had more impact on the TinySleepNet performance than the subject conditions.
When all three characteristics differed, the $r_{i,j}$ of TinySleepNet was the highest at 29.4\%.
According to these results, the impact of each data characteristic when directly applying the pre-trained TinySleepNet model to a target dataset can be ranked in ascending order as follows: subject conditions, recording channels, and recording environments.

For U-Time, the $r_{i,j}$ showed different patterns than TinySleepNet.
When the recording environments were the same, either the differences in recording channels or subject conditions (5.3\% and 2.7\%) had less $r_{i,j}$ than when they were the same (7.4\%).
When the environments differed, the $r_{i,j}$ rose to more than 14\% in all configurations.
Interestingly, the $r_{i,j}$ when the subject conditions were the same (30.2\% and 31.6\%) was higher than when they were different (14.6\% and 25.2\%).
The highest $r_{i,j}$ of U-Time was 31.5\% when the recording channels and environments were different but subject conditions were similar.
According to these results, the impact of each data characteristic when directly applying the pre-trained U-Time model to a target dataset can be ranked in ascending order as follows: recording channels, subject conditions, and recording environments.
They also suggest that applying the pre-trained U-Time model to a target dataset would be better when the subject conditions differ.

According to our results, the recording environments had the highest impact on the model performance by more than 17\% and 14\% for TinySleepNet and U-Time, respectively, and should be minimized when employing a model in practical applications.
The recording channels and subject conditions impacted TinySleepNet and U-Time differently and should be quantified when considering other deep learning models.

\begin{figure*}[!t]
  \begin{subfigure}[b]{0.81\linewidth}
    \centerline{\includegraphics[width=\linewidth]{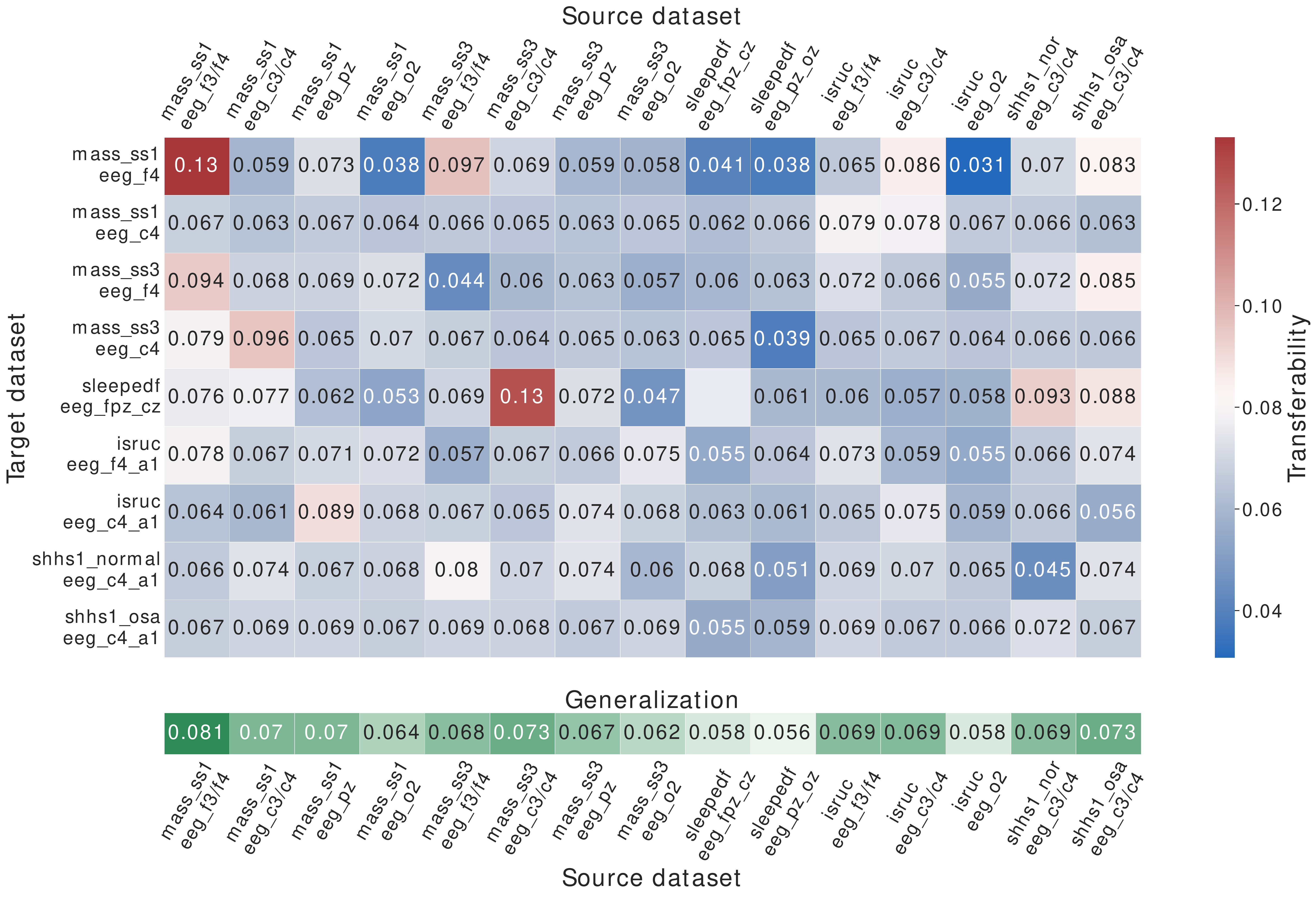}}
    \caption{$W$ from TinySleepNet.}
    \label{fig:tf_mat_30s_tinysleepnet}
  \end{subfigure}
  \centering
  \vfill 
  \begin{subfigure}[b]{0.80\linewidth}
    \centerline{\includegraphics[width=\linewidth]{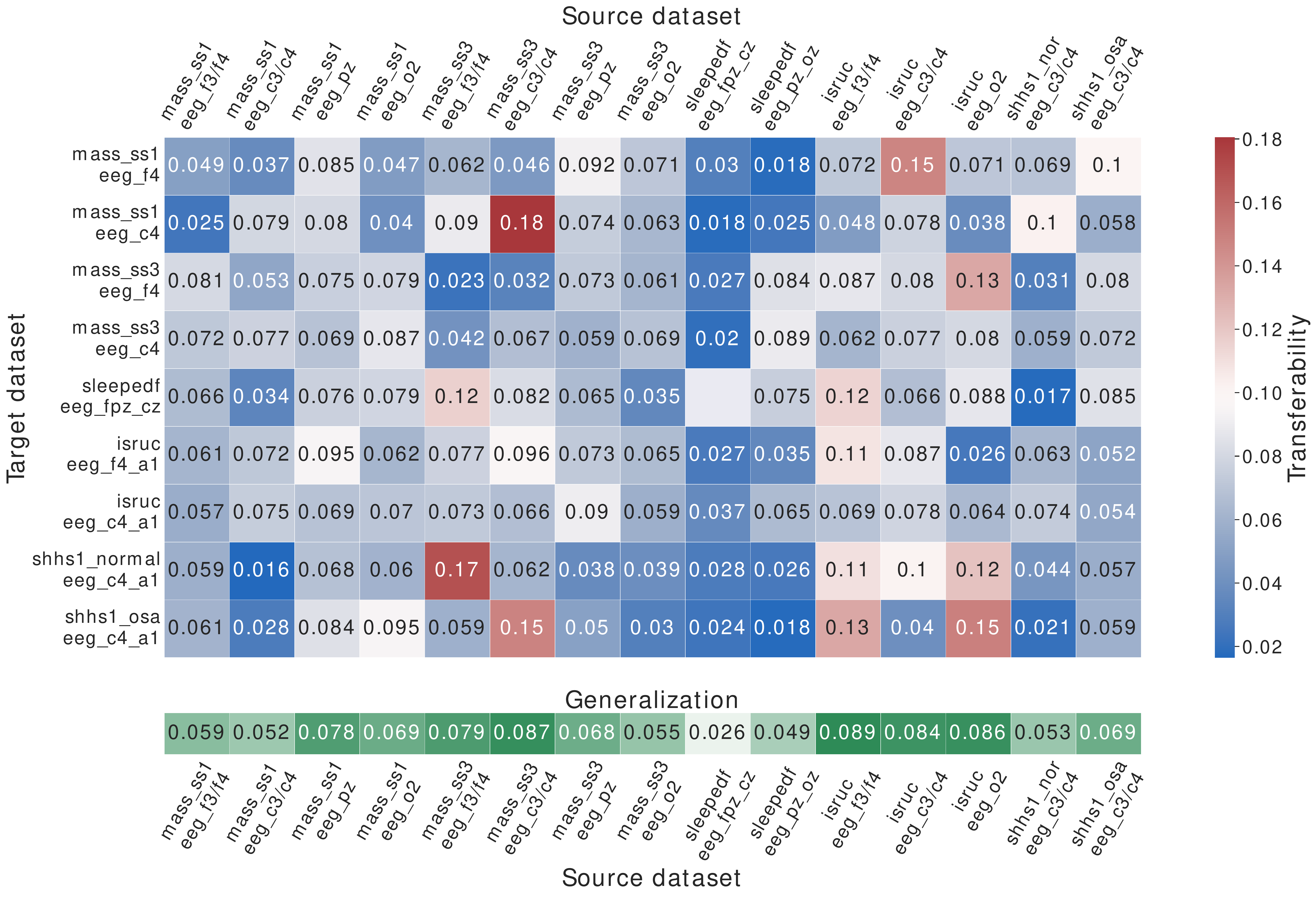}}
    \caption{$W$ from U-Time.}
    \label{fig:tf_mat_30s_utime}
  \end{subfigure}
  \centering
  \captionsetup{width=\linewidth}
  \caption{Transferability matrices ($W$) from TinySleepNet (a) and U-Time (b) models.
  Each element $w_{i,j}$ represents the transferability or the usefulness of a source $d_j$ to a target $d_i$.
  For each row (target), the higher degree of transferability indicates that such a dataset is a better source dataset than the others when transferring to the target dataset.
  The generalization vectors at the bottom of the matrices are the average of each column (source) of $W$, representing the usefulness of each dataset compared to the other datasets.}
  \label{fig:result_transferability}
\end{figure*}

\subsection{Transferability}
To answer RQ2, two transferability matrices, $W$, for TinySleepNet (Fig.~\ref{fig:tf_mat_30s_tinysleepnet}) and U-Time (Fig.~\ref{fig:tf_mat_30s_utime}) were computed.
We found that even though the recording channels, recording environments, and subject conditions between the sources and the target were different, the model still performed as well as other more similar sources after fine-tuning.
For instance, Fig.~\ref{fig:tf_mat_30s_tinysleepnet} shows the EEG-C4 from MASS-SS3 (0.13) had a higher transferability to the EEG-Fpz-Cz from Sleep-EDF-SC than the EEG-Pz-Oz from the same dataset (0.061), even though they were from the different recording environments.
Similarly, Fig.~\ref{fig:tf_mat_30s_utime} shows that EEG-C4 from MASS-SS3 (0.15) had a higher transferability to EEG-C4 from SHHS1-OSA than EEG-C3 from the same dataset, even though they were from different environments and subject conditions.
This suggests that the pre-trained TinySleepNet and U-Net models generalize well to new datasets.

To determine which pre-trained model had a higher degree of generalization or which dataset was most useful among all considered datasets, we compared the average of each source (column) of $W$ (the vector at the bottom of each transferability matrix in Fig.~\ref{fig:result_transferability}).
For TinySleepNet, the EEG-F3/F4 from MASS-SS1 (0.081) was the most useful transfer source, while the EEG-Pz-Oz from Sleep-EDF-SC (0.056) was the least useful one.
For U-Time, EEG F3/F4 from ISRUC-SG1 (0.089) was the most useful source, while EEG-Fpz-Cz from Sleep-EDF-SC (0.026) was the least useful one.
On average, MASS-SS1 and ISRUC-SG1 were the most useful transfer sources for TinySleepNet and U-Time, while Sleep-EDF-SC was the least useful transfer source for both models.

In addition, we investigated the EEG channels from which areas of the brain would be more useful compared to the other areas.
Interestingly, we found that the source datasets with the EEG channels from the frontal (F) and the central (C) areas of the brain were more useful than the parietal (P) and the occipital (O) for TinySleepNet.
However, we have yet to find any consistent patterns for which EEG channel would be more useful compared to the others for U-Time.

\section{Discussion}

We presented a novel method of quantifying the impact that various data characteristics have on the sleep stage scoring performance and the transferability of a deep learning model that scores sleep stages based on raw single-channel EEG.
The results from two deep learning models, TinySleepNet and U-Time, demonstrate that the pre-trained models did not perform as well on a target dataset with different characteristics than with similar ones.
The recording environments play an essential role when directly applying the pre-trained models to a target dataset (direct transfer).
Different environments would amplify the impact of differences in recording channels and subject conditions, increasing the $r_{i,j}$ by more than 17\% and 14\% for TinySleepNet and U-Time.
Such a difference is expected when applying the model trained on the public sleep datasets to new devices (e.g., wearable devices).
Thus, when the sleep annotations of the target datasets are unavailable, the models pre-trained on the datasets with similar recording environments are recommended for both models.
This study enables us to estimate in advance \textit{how much} performance degradation one can expect from \textit{each different characteristic} when directly applying a pre-trained model to a particular dataset (RQ1).

When a labeled target dataset is available, transfer learning can be used and the results show that the model fine-tuning with any source dataset helps improve the performance on the target dataset (Fine-tuned Transfer (FT) > From Scratch (FS)).
On average, the most useful transfer sources for TinySleepNet and U-Time were the MASS-SS1 and ISRUC-SG1 datasets, while the least useful source for both models was the Sleep-EDF-SC dataset.
Even though the most useful datasets were different for both models, we noticed that both datasets have one thing in common: the high percentage of samples from the N1 stage (the rarest class) relative to the other stages: 13.9\% for MASS-SS1 and 12.7\% for ISRUC-SG1 compared to the other datasets with less than 9\%.
These results suggest that the recommended transfer sources for both models are the datasets with a high percentage of N1 samples relative to the other stages (RQ2).
Further experiments on a larger number of sleep datasets with various data characteristics are still necessary to confirm our findings.
Moreover, we found that the EEG channels from the frontal and the central brain areas were more useful than the parietal and the occipital for TinySleepNet (RQ2).
However, we have yet to find any consistent patterns of the useful brain areas for U-Time.

The proposed method can also be considered a new method to determine how well newly proposed deep learning models generalize to new datasets, which may have different characteristics.
This is a typical question when one would like to employ a model for remote sleep monitoring in the home environment.
Thus, it is useful to know which characteristics would affect the performance of a particular model the most and what source datasets can be used to alleviate performance degradation.
This method introduces a new approach to evaluate a newly proposed deep learning model to show its generalizability across datasets with different characteristics.

Even though this study presents an interesting research direction to understand the impact of various data characteristics on the transferability of deep learning models, we have only scratched the surface.
Several limitations are worth mentioning.
First, even though we evaluated our approach on many public sleep datasets with different characteristics and repeated the experiments several times for reliability, the results were still model- and data-specific.
The answers to the two research questions may vary depending on the choices of deep learning models and sleep datasets.
As the datasets considered in this study represent only a subset of the real-world datasets, there may be other characteristics that affect sleep stage scoring performance.
Second, our approach relies on the performance of the fine-tuned models to quantify transferability.
This means we can only estimate the transferability of new datasets when the sleep annotations are available.
Third, the models pre-trained on all source datasets are not publicly available and are assumed to be trained by researchers.

\section{Conclusion}
This paper presents a new way of quantifying the impact of various characteristics of sleep datasets on the direct applicability of pre-trained models and their transferability (or usefulness) to new target problems.
We examined the applicability of pre-trained models by conducting experiments using two deep learning models, TinySleepNet and U-Time, and six public sleep datasets. The results showed that the higher the degree of dissimilarity of data characteristics between the source and target, the lower the model performance on the target problem.
The recording environment had the most significant impact on model performance when directly applying the pre-trained model to a target problem.
When a labeled target dataset is available, transfer learning can be used, and our results show that model fine-tuning helps improve the model performance on the target dataset, regardless of the differences in considered characteristics.
In addition, since our method involves training and validation across various datasets with different characteristics, it provides a new evaluation perspective that looks at the generalization of a newly proposed model across datasets.

In the future, we plan to extend our approach to predict model transferability without running intensive transfer learning across sleep datasets. This will help reduce the cost of pre-training models on source datasets and fine-tuning them to target datasets. By developing a more efficient and cost-effective method for predicting transferability, we hope to enable researchers and practitioners to optimize the performance of pre-trained models on new datasets with different characteristics, even when labeled target datasets are limited or not readily available.

\section*{Acknowledgement}
This research project is supported by Mahidol University (Basic Research Fund: fiscal year 2021).

\section*{Declarations of interest}
None.

\section*{Conflict of interest statement}
None.




\bibliographystyle{cas-model2-names}

\bibliography{main}

\end{document}